\documentclass[12pt]{iopart}

\usepackage{epsfig}
\newcommand{\putfig}[5]{
\begin{figure}[hptb]
         \begin{center}
             \leavevmode
             \psfig{file=#1,height=#2,width=#3}
         \end{center}
         \vspace{-2ex}
         \caption{#4}
         \label{#5}
\end{figure}
}

\usepackage{iopams,epsfig}
\newcommand{\bea}{\begin{eqnarray}}
\newcommand{\eea}{\end{eqnarray}}
\newcommand{\ba}{\begin{array}}
\newcommand{\ea}{\end{array}}
\newcommand{\half}{\frac{1}{2}}

\newcommand{\be}{\begin{equation}}

\newcommand{\ee}{\end{equation}}
\newcommand{\bt}{\begin{teo}}
\newcommand{\et}{\end{teo}}
\newcommand{\ep}{\epsilon}
\newcommand{\s}{\sigma}
\newcommand{\la}{\lambda}
\newcommand{\om}{\omega}
\newcommand{\ii}{{\rm i}}
\newcommand{\imag}{{\rm i}}
\newcommand{\ek}{{e^{\frac{r(\lambda_1)}\epsilon}}}
\newcommand{\ekd}{{e^{2\frac{ r(\lambda_1)}\epsilon}}}
\newcommand{\bn}{B_0}
\newcommand{\bk}{B_1}
\newcommand{\an}{A_0}
\newcommand{\ak}{A_1}
\newcommand{\on}{\omega_0}
\newcommand{\ok}{\omega_1}
\newcommand{\onp}{\omega'(\la_0)}
\newcommand{\okp}{\omega'(\la_1)}
\newcommand{\si}{\sin \left(\frac {s_1}\epsilon \right)}
\newcommand{\co}{\cos \left(\frac {s_1}\epsilon \right)}
\newcommand{\siq}{\sin^2 \left(\frac {s_1}\epsilon \right)}
\newcommand{\coq}{\cos^2 \left(\frac {s_1}\epsilon \right)}

\begin{document}

\epsfig{file=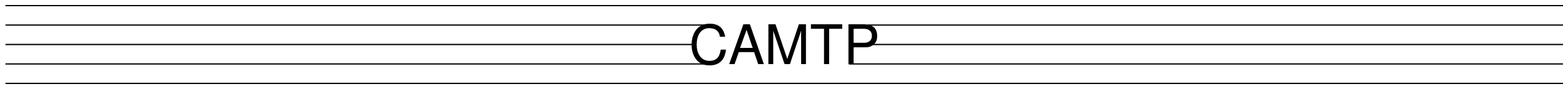,height=8mm,width=\textwidth}

\title[Energy evolution in time-dependent harmonic
oscillator]
{Energy evolution in time-dependent harmonic oscillator }

\author{Marko Robnik$^*$ and Valery G. Romanovski}

\address{CAMTP - Center for Applied Mathematics and Theoretical Physics,
University of Maribor, Krekova 2, SI-2000 Maribor, Slovenia}
\address{$^*$ ATR Advanced Telecommunications Research Institute International,
2-2-2 Hikaridai, Seika-cho, Soraku-gun, Kyoto 619-0288, Japan}


\eads{Robnik@uni-mb.si, Valery.Romanovsky@uni-mb.si}

\begin{abstract}
The theory of adiabatic invariants has a long history, and
very important implications and applications in many different
branches of physics, classically and quantally, but is rarely
founded on rigorous results. Here we treat the general time-dependent
one-dimensional harmonic oscillator, whose Newton equation
$\ddot{q} + \omega^2(t) q=0$ cannot be solved in general.
We follow the time-evolution of an initial ensemble of 
phase points with sharply defined energy $E_0$ at time $t=0$
and calculate rigorously the distribution of energy $E_1$ after time $t=T$,
which is fully (all moments, including the variance $\mu^2$) 
determined by the first moment $\bar{E_1}$. For example, 
$\mu^2 = E_0^2 [(\bar{E_1}/E_0)^2 - (\omega (T)/\omega (0))^2]/2$,
and all higher even moments are powers of $\mu^2$, whilst the odd ones 
vanish identically. This distribution function does not depend 
on any further details of the function $\omega (t)$ and is 
in this sense universal.
In ideal adiabaticity $\bar{E_1} = \omega(T) E_0/\omega(0)$,
and the variance $\mu^2$ is zero, whilst for finite $T$ we
calculate $\bar{E_1}$, and $\mu^2$ for the general case
using exact WKB-theory to all orders. We prove that if $\omega (t)$ 
is of class ${\cal C}^{m}$ (all derivatives up to and
including the order $m$ are continuous)
$\mu \propto T^{-(m+1)}$, whilst for class ${\cal C}^{\infty}$ it 
is known to be exponential $\mu \propto \exp (-\alpha T)$. 

\end{abstract}


Published in {\em Open Systems \& Information Dynamics} 
{\bf 13} No.2 p.197-222 (2006) 

\section{Introduction}

In time-independent (autonomous) Hamiltonian systems the total energy
of the system is conserved by construction, i.e. due to the Hamilton
equations of motion, and the Liouville theorem applies, because
the phase flow velocity vector field has vanishing divergence. 
In time-dependent (nonautonomous)
Hamilton systems the total energy is not conserved, whilst the
Liouville theorem still applies, the phase space volume is preserved
under the phase flow. If (some parameter of) the Hamilton function 
varies in time, the energy of the system generally also changes.
But, if the changing of the parameter
is very slow, on the typical time scale $T$,
there might be a quantity $I$, a function of the said parameter, 
of the energy $E$ and of other dynamical quantities, which is 
approximately conserved. 
It might be even exactly conserved if $T\rightarrow \infty$,
i.e. if the variation is infinitely slow, to which case we refer as the
ideal adiabatic variation. 
Such a conserved quantity is called {\em adiabatic invariant}, 
and it plays an important role in the dynamical analysis of a long-time
evolution of nonautonomous Hamilton systems. 

The theory of adiabatic invariants is aimed at finding the adiabatic
invariants $I$ and analyzing the error of its preservation at
finite $T$. Namely, the statement of exactness of $I$
is asymptotic in the sense that the conservation is exact in
the limit $T \rightarrow \infty$, whilst for finite $T$ we 
see the deviation $\Delta I = I_{f}-I_{i}$ of final value of $I_{f}$ 
from its initial value $I_{i}$ and would like to calculate
$\Delta I$. Thus for finite $T$ the final values of $I$ will have
some distribution with nonvanishing variance.

Here we just mention, as an example, that for one-dimensional 
harmonic oscillator it is known since Lorentz and Einstein \cite{Einstein}
that the adiabatic invariant for $T=\infty$ is $I=E/\omega$, which is
the ratio of the total energy $E=E(t)$ and the frequency of the oscillator
$\omega (t)$, both being a function of time. Of course, $2\pi I$ is
exactly the area in the phase plane $(q,p)$ enclosed by the energy contour
of constant $E$. 
A general introductory account of the theory of adiabatic invariants
can be found in \cite{Rob}
and references therein, especially \cite{LL}, \cite{Rein}.
Another quite extensive excellent review is by Henrard \cite{Henrard},
which includes rather complete list of relevant references to the
original papers.

However, in the literature this $I$ and $\Delta I$ are 
not even precisely defined.
It turns out that $I$ must be generally considered as a function 
of the initial conditions, and then it turns out that it is conserved
for some initial conditions but not for some others.
As a consequence of that there is a considerable confusion about
its meaning. Let us just mention the case of periodic 
parametric resonance (see section 5), with otherwise 
arbitrarily slowly changing
 $\omega (t)$, in one-dimensional harmonic oscillator,
where the total energy of the system can grow
indefinitely for certain (almost all) initial conditions, 
and since $\omega (t)$ 
is bounded, $I= E(t)/\omega (t)$ simply cannot be conserved
for the said initial conditions for all times, but only for
sufficiently small times.
In this work we consider $I$ as function of the initial
conditions, and give a precise meaning to these and similar statements.

Therefore to be on rigorous side we must carefully define what we mean by
$I$ and $\Delta I$. This can be done by considering an ensemble of initial
conditions at time $t=0$ just before the adiabatic process starts.
Of course, there is a vast freedom in choosing such ensembles.
Let us consider the one degree of freedom systems, which is the 
topic of this paper. 
If the initial conditions are on a closed contour ${\cal K}_0$ 
in the phase space
at time $t=0$, then at the end of the adiabatic process at time $t=T$
they are also on a closed contour ${\cal K}_T$ which in general 
is different from ${\cal K}_0$, but due to the Liouville theorem
the area inside the contour is constant for all $T$. 
If ${\cal K}_0$ is a contour of constant energy at time $t=0$,
then ${\cal K}_T$ generally is not a contour of constant
energy of the Hamiltonian at time $T$. The final energies of the
system, depending on the initial conditions, are spread and
thus distributed between some minimal and maximal energy, $E_{min}$
and $E_{max}$, respectively.
In the said case of periodic $\omega (t)$ and parameter resonance 
in a harmonic oscillator
the contour  ${\cal K}_T$ is squeezed in one direction
and expanded in the other one (the transformation is a linear map),
and this contraction and expansion is exponential in time. 
Therefore, $I = E(T)/\omega (T)$ can not be conserved; see section 5.

Thus we must always study $I$ as a function of initial conditions,
by looking at the ensembles of initial conditions. There is a vast
freedom in choosing such ensembles, but usually we do not
know very much about the system except e.g. just the energy.

Therefore in an integrable conservative Hamiltonian system the most natural 
and the most important 
choice is taking as the initial ensemble all phase points 
uniformly distributed on the initial $N$-torus, i.e.  uniform w.r.t. the angle
variables. We call it {\em uniform canonical ensemble of initial conditions}.
Such an ensemble has a sharply defined initial energy $E_0$.
Then we let the system evolve in time, not necessarily slowly,
and calculate the probability distribution $P(E_1)$ of the final energy $E_1$,
or of other dynamical quantities.  Typically $E_1$ is distributed
on an interval $(E_{min}, E_{max})$, and $P(E_1)$ is universal there,
as it does not depend on any further properties of $\omega (t)$
except for $\bar{E_1}$. More general ensembles of
initial energies $w(E_0)$ can be described in terms of the 
uniform canonical ensembles, as explained in section 6.

To describe $P(E_1)$ is in general a difficult problem, 
but  in this work we confine
ourselves to the one-dimensional general time-dependent harmonic oscillator,
so $N=1$, described by the Newton equation  

\be \label{eq:Newton}
\ddot{q} + \omega^2 (t) q =0
\ee
and work out rigorously $P(E_1)$. Given the general 
dependence of the oscillator's frequency $\omega(t)$ on time $t$ the
calculation of $q(t)$ is already a very difficult, in fact unsolvable, 
problem.  In the sense of mathematical physics  (\ref{eq:Newton})
is exactly equivalent to the one-dimensional
stationary Schr\"odinger equation: the coordinate $q$ appears
instead of the probability amplitude $\psi$, time $t$ appears instead of the
coordinate $x$ and $\omega^2(t)$ plays the role of $E-V(x)$ = energy --
potential. If $E$ is greater than any local maximum of $V(x)$
then the scattering problem is equivalent to our 1D harmonic oscillator
problem.
In this paper we solve the above stated  problem for the 
general one-dimensional harmonic oscillator which is a follow up
paper of our Letter \cite{RR2006}.

In performing our analysis, we shall answer the questions
as to when is $I=E(T)/\om (T)$ conserved, and if it is not conserved,
what is the spread or variance   $\mu^2$ of the energy, 
and the higher moments etc. Then (the
not sharply defined) $\Delta I$ in the literature is proportional
to $\mu$. After performing the exact analysis, we provide a powerful
technique based on the WKB method \cite{RR2000}
to calculate $\mu^2$, and show that
it gives exact leading asymptotic terms when $T\rightarrow \infty$,
and moreover, generally we can do the expansion to all orders, exactly.
We treat several exactly solvable cases, and compare them with
the WKB results, and finally prove the theorem as for how $\mu^2$
behaves when $\omega (t)$ is of class ${\cal C}^m$, which means
having $m$ continuous derivatives.
 
We give a brief historical review of contributions
to this field.  After Einstein \cite{Einstein}, 
Kulsrud \cite{Kulsrud}  was the first to show, using a WKB-type method,
that for a finite $T$, $I$ is preserved to
all orders, for harmonic oscillator, if all derivatives of $\omega$
vanish at the beginning and at the end of the time interval, 
whilst if there is
a discontinuity in one of the derivatives he estimated 
$\Delta I$ but did not give our explicit general expressions 
(37) and (38). Hertweck and Schl\"uter \cite{Hertweck}
did  the same thing  independently
 for a charged particle in slowly
varying magnetic field for infinite time domain.
Kruskal, as reported in \cite{Gardner}, and  Lenard \cite{Lenard} studied
more general systems, whilst Gardner \cite{Gardner} used the classical
Hamiltonian perturbation theory. Courant and Snyder \cite{Courant} have
studied the stability of synchrotron
and analyzed $I$ employing the transition matrix. 
The interest then shifted to the infinite time domain.
Littlewood \cite{Littlewood}
showed for the harmonic oscillator that if $\omega(t)$ is
an analytic function, $I$ is preserved to all orders of the 
{\em adiabatic parameter} $\ep=1/T$. 
Kruskal \cite{Kruskal2} developed the asymptotic theory
of Hamiltonian and other systems with all solutions nearly periodic.
Lewis \cite{Lewis}, using the Kruskal's method, discovered a connection 
between $I$ of the 1-dim harmonic oscillator and
another nonlinear differential equation.
Later on Symon \cite{Symon} used the Lewis'es
results to calculate the (canonical) ensemble average
of the $I$ and its variance, which is the analogue
of our $\bar{E_1}$ and $\mu^2$. Finally, Knorr and Pfirsch
\cite{Knorr} proved $\Delta I \propto \exp (-const/\ep)$.
Meyer \cite{Meyer1}, \cite{Meyer2} relaxed some conditions
and calculated the constant $const$. Exponential preservation
of $I$ for an analytic $\omega$ on $(-\infty, +\infty)$
with constant limits at $t \rightarrow
\pm \infty$, is thus well established \cite{LL}.

\section{Transition map and general exact considerations}

We begin by defining the system by giving its Hamilton function 
$H = H(q,p,t)$, whose numerical value $E(t)$ at time $t$ is precisely the total
energy of the system at time $t$, and for the general time-dependent
one-dimensional harmonic oscillator this is

\be \label{eq:Hamilton}
H=\frac {p^2}{2M}+\frac 12 M \omega^2(t)q^2,
\ee
where $q,p,M,\omega$ are the coordinate, the momentum, the mass and
the frequency of the linear oscillator, respectively. 
The dynamics is linear in $q,p$, as described by (\ref{eq:Newton}),
but nonlinear as a function of $\omega (t)$ and therefore is subject to
the nonlinear dynamical analysis. By using the index $0$ and $1$ we denote the
initial ($t=t_0$) and final $(t=t_1)$ value of the variables, 
and by $T=t_1-t_0$ we denote  the length of the time interval of 
changing the parameters of the system.

We consider the phase flow  map (we shall call it transition map)

\be
\Phi:\left( 
\begin{array}{c}
q_0\\ p_0
\end{array}
 \right)\mapsto  \left( 
\begin{array}{c}q_1\\ p_1
\end{array}
 \right).
\ee
 Because equations of motion are linear in $q$ and $p$, and since
the system is Hamiltonian,
$\Phi$ is a linear  area preserving map, that is, 

\be \label{eq:transitionmap}
\Phi=\left(
\begin{array}{cc}
 a & b\\
  c & d
\end{array}
\right),
\ee
with  $\det(\Phi)=ad-bc=1$.
Let $E_0=H(q_0,p_0,t=t_0)$ be the initial energy and 
$E_1=H(q_1,p_1,t=t_1)$ be the final energy, that is,

\begin{equation} \label{eq:em1}
E_1=\frac 12 \left(\frac { (cq_0+d p_0)^2}M+M \omega^2_1(a q_0+b p_0)^2\right).
\end{equation}
Introducing the new coordinates, namely the action $I=E/\omega$ and
the angle $\phi$, 

\be
 q_0=\sqrt{\frac{2 E_0}{M\omega_0^2}}\cos \phi,\ 
 p_0=\sqrt{{2M  E_0}}\sin \phi
\ee
 from (\ref{eq:em1}) we obtain
\be \label{eq:ee1}
E_1=E_0(\alpha \cos ^2 \phi+\beta \sin^2 \phi +\gamma \sin 2\phi),
\ee
where 

\be \label{eq:abg}
\alpha  = \frac{c^2}{M^2 \omega_0^2}+a^2 \frac{\omega_1^2}{\omega_0^2},\;\;\;\;
 \beta = d^2+ \omega_1^2 M^2 b^2,\;\;\;\;
\gamma   = \frac{c d}{M \omega_0} + ab M \frac{\omega_1^2}{\omega_0}.
\ee
Given the uniform probability distribution of initial angles $\phi$
equal to $1/(2\pi)$, which defines our initial {\em uniform canonical
ensemble} at time $t=t_0$, we can now calculate the averages. Thus

\be \label{eq:bare1}
\bar E_1=\frac 1{2\pi}\oint E_1d\phi=\frac {E_0}2 (\alpha+\beta).
\ee 
That yields $E_1-\bar E_1=E_0 (\delta \cos 2\phi+\gamma \sin 2 \phi)$
and 

\be \label{eq:m2}
\mu^2 = 
\overline{(E_1-\bar E_1)^2}=\frac {E_0^2}2 \left(\delta^2+\gamma^2 \right),
\ee
where we have denoted $\delta=(\alpha-\beta)/2$. 
It follows from (\ref{eq:abg}), (\ref{eq:bare1})  that we can write 
(\ref{eq:m2})  also in the form

\be \label{eq:sigma}
\mu^2 =
\overline{(E_1-\bar E_1)^2}=\frac {E_0^2}2 \left[\left( \frac{\bar E_1}{E_0} 
 \right)^2 - \left(\frac{\om_1}{\om_0}\right)^2 \right].
\ee
As we shall see (see subsection 4.2), in an ideal adiabatic process 
$\mu=0$, and therefore $E_1= \bar{E_1} = \om_1E_0/\om_0$,
and consequently $P(E_1)$ is a delta function, 

\be \label{eq:idealadia}
P(E_1) = \delta (E_1 - \om_1E_0/\om_0).
\ee

Now we calculate higher moments of $P(E_1)$.
Let us show that in general for arbitrary positive integer $m$ 

\be \label{eq:emodd}
\overline{(E_{1}-\bar E_{1})^{2m-1}}=0
\ee
and

\be\label{eq:emeven}
\overline{(E_{1}-\bar E_{1})^{2m}}=\frac {(2m -1)!!}{ m!}
\left( \overline{(E_{1}-\bar E_{1})^2}\right)^m.
\ee
It is easy to check the correctness of (\ref{eq:emodd}), so we prove only 
(\ref{eq:emeven}). Indeed,

\be\label{eq:esc}
\overline{(E_{1}-\bar E_{1})^{2m}}=E_0 ^{2m}
\overline{(\gamma \sin 2\phi +\delta \cos 2 \phi)^{2m} }=
\ee
$$
E_0^{2m} \sum_{k=0}^{2m} {2m \choose k}
\overline{( \gamma^{2m-k} \delta^k  \sin^{2m -k} 2 \phi \cos^k 2\phi )  }.
$$
Note that for odd  $k=2l-1$ 

\be
\overline{\sin^{2m -k} 2 \phi \cos^k 2\phi}=0.
\ee
Therefore we can rewrite (\ref{eq:esc}) in the form 

\be \label{eq:e16}
\overline{(E_{1}-\bar E_{1})^{2m}}=E_0^{2m} \sum_{l=0}^{m} {2m \choose 2l}
\overline{( \gamma^{2m-2l} \delta^{2l}  \sin^{2m -2l} 
2 \phi \cos^{2l} 2\phi )  }.
\ee
Using formula 2.512 of \cite{GR} we obtain 

\be
\frac 1{2 \pi}
\int_0^{2\pi}\sin^{ 2 (m-l)} 2\phi \cos^{2l} 2 \phi\, d\phi=
\frac{ (2l-1)!! (2m -2l-1)!!}{2^m m!}.
\ee
The substitution of this value into (\ref{eq:e16})
yields

\bea
\overline{(E_{1}-\bar E_{1})^{2m}}=E_0^{2m} \sum_{l=0}^{m} {2m \choose 2l}
{\left( \gamma^{2m-2l} \delta^{2l} \frac { (2l-1)!! (2m -2l-1)!!}{2^m m!} \right)  }\\
=E_0^{2m} \sum_{l=0}^{m} 
{\left( \gamma^{2m-2l} \delta^{2l} \frac {  (2m )!}{2^m m! (2l)!!(2m-2l)!! } \right)  } \\
= E_0^{2m}  \frac {  (2m -1)!!}{2^m m!  }   \sum_{l=0}^{m}
{m \choose l} 
 \gamma^{2m-2l} \delta^{2l} \\
  = \frac {  (2m -1)!!}{2^m m!  } \left(\gamma^2+\delta^2\right)^m E_0^{2m}=
\frac {  (2m -1)!!}{ m!  } 
\left( \overline{(E_{1}-\bar E_{1})^2}\right)^m, 
\eea
that is, the formula  (\ref{eq:emeven}) holds.
Thus $2m$-th moment of $P(E_1)$ is equal to $(2m-1)!! \mu^{2m}/m!$,
and therefore, indeed, all moments of $P(E_1)$ are uniquely determined 
by the first moment $\bar{E_1}$.

We now mention that the final energy distribution function written down as

\be \label{eq:Edistribution}
P(E_1)=\frac 1{2\pi} \sum_{j=1}^4\left| \frac{d\phi}{dE_1} 
\right|_{\phi=\phi_j(E_1)}
\ee
cannot be calculated analytically in a closed form in any useful way, 
because it boils down to finding the roots of a quartic polynomial, so
we do not try to do that here, although numerically it  shows
interesting aspects (see figure 3). 
Obviously, $P(E_1)$ is in this sense
universal, because it depends only on the average final energy $\bar{E_1}$ and
the ratio $\ok/\on$ of the final and initial frequencies, and 
does not depend otherwise on any details of $\om (t)$.
It has a finite interval as its support, between
the lower limit $E_{min}$ and the upper limit $E_{max}$, it is an even 
function there w.r.t. the mean value $\bar{E_1} = (E_{min} + E_{max})/2$,
and from  $E_1-\bar E_1=E_0 (\delta \cos 2\phi+\gamma \sin 2 \phi)$
it is easy to show that $E_{min} = \bar{E_1} - \mu\sqrt{2}$
and $E_{max} = \bar{E_1} + \mu \sqrt{2}$, so that 
$E_{max} - E_{min} = \mu\sqrt{8} =2E_0\sqrt{\gamma^2+\delta^2}$.
At both extreme values it has an integrable 
singularity of the type $1/\sqrt{x}$.
This singularity
stems from a projection of the final ensemble at $t_1=T$ onto the curves of
constant final energies $E_1$ of $H(q,p,t_1)$.
 In between
for every value of $E_1 = const = E_1(\phi)$, this equation has four solutions,
namely $\phi_1,\phi_2,\phi_3,\phi_4$, and thus we have to sum up
all four contributions in the general formula (\ref{eq:Edistribution}).
 Of course, all that we say
here for the distribution of energies $E_1$ holds true also for the 
final action, the adiabatic invariant  $I_1=E_1/\om_1$. It is perhaps
worthwhile to mention that the moments of our distribution according to
(\ref{eq:emeven}) grow as $2^m/\sqrt{\pi m}$, whilst e.g. in the Gaussian
distribution they grow much faster, namely as $2^m\Gamma (m+1/2)/\sqrt{\pi}$,
where $\Gamma (x)$ denotes the gamma function.

Expression (\ref{eq:sigma}) is positive definite by definition and this
leads to the first interesting conclusion: In full generality 
(no restrictions on the function $\omega (t)$!) we have always
$\bar{E_1} \ge E_0\omega_1/\omega_0$ and therefore the final value
of the adiabatic invariant (for the average energy!) 
$\bar{I_1}=\bar{E_1}/\omega_1$ is always greater
or equal to the initial value $I_0=E_0/\omega_0$. In other words,
the value of the adiabatic invariant at the mean value of the energy
 never decreases, which is
a kind of irreversibility statement. Moreover, it is conserved only
for infinitely slow processes $T=\infty$, which is an ideal adiabatic 
process, for which $\mu=0$. See section 4.3.
For periodic processes $\omega_1=\omega_0$ we see that 
always $\bar{E_1} \ge E_0$, so the mean energy never decreases.
See section 5.
The other extreme opposite to $T=\infty$ is the instantaneous ($T=0$) 
jump where $\omega_0$
switches to $\omega_1$ discontinuously, whilst $q$ and $p$ remain
continuous, and this results in $a=d=1$ and $b=c=0$, and then we find

\be \label{eq:jump}
\bar{E_1} = \frac{E_0}{2} (\frac{\omega_1^2}{\omega_0^2} + 1),\;\;\;\;\;
\mu^2 = \frac{E_0^2}{8} \left[ \frac{\omega_1^2}{\omega_0^2} -1\right]^2.
\ee
Below we shall treat the special case with $\omega_1^2 = 2\omega_0^2$,
and thus will find $\mu^2/E_0^2 = 1/8 = 0.125$.

Our general study now focuses on the calculation of the transition
map (\ref{eq:transitionmap}), namely its matrix elements $a,b,c,d$.
Starting from the Hamilton function (\ref{eq:Hamilton})  
and its Newton equation (\ref{eq:Newton})
we consider two  linearly  independent solutions $\psi_1(t)$ 
and $\psi_2(t)$ and introduce the matrix

\be \label{eq:phig}
\Psi(t)=\left(
\ba{cc}
 \psi_1(t) & \psi_2(t)\\
  M\dot{\psi}_1(t) & M\dot{\psi}_2(t)
\ea
\right).
\ee
Consider  a solution $\hat q(t)$ of (\ref{eq:Newton})
such that

\be \label{eq:qp0}
\hat q(t_0)=q_0,\quad \dot { \hat q}(t_0)=p_0/M.
\ee
Because $\psi_1$ and $\psi_2$ are linearly independent, we can look for  
 $\hat q(t)$ in the form

\be
\hat q(t)=A \psi_1(t) + B \psi_2(t). 
\ee
Then $A$ and $B$ are determined by 

\be \label{eq:AB}
\left( 
\begin{array}{c}
A\\ B
\end{array}
 \right)= \Psi^{-1}(t_0) \left( 
\begin{array}{c}q_0\\ p_0
\end{array}
 \right).
\ee
Let  $q_1=\hat q(t_1),\quad  p_1=M \dot {\hat q}(t_1)$.
Then from  (\ref{eq:qp0})--(\ref{eq:AB}) we  see that

\be \label{eq:p1q1}
\left( 
\begin{array}{c}
q_1\\ p_1
\end{array}
 \right)= \Psi(t_1)  \Psi^{-1}(t_0) \left( 
\begin{array}{c}q_0\\ p_0
\end{array}
 \right).
\ee
We recognize the matrix on the right-hand side of (\ref{eq:p1q1}) 
as the transition map $\Phi$, that is,

\be \label{eq:phiT}
\Phi=\left(
\ba{cc}
 a & b\\
  c & d
\ea
\right)= \Psi(t_1)  \Psi^{-1}(t_0).
\ee

\section{Some exactly solvable special cases}

The theory of adiabatic invariants is rarely founded on rigorous results,
therefore the study of exactly solvable cases is of fundamental
importance, namely we can use them to test various analytic approximations
and also the accuracy of numerical calculations. Because the ideal adiabatic 
processes are infinitely slow and refer to the limit $T\rightarrow \infty$ 
we must deal necessarily with the asymptotic behaviour of dynamical systems,
which is difficult to approximate, because all simple minded perturbational
and other approximation techniques fail (break down) after some finite
time, and so typically give wrong predictions for the asymptotic behaviour
in the limit $T\rightarrow \infty$. The same difficulty occurs in
numerical calculations. We shall see that the WKB methods
\cite{RR2000} can be successfully applied. 
In this section we deal with three different
models for the function $\omega^2 (t)$.

\subsection{The linear model: class ${\cal C}^0$}

We assume that function $\omega^2(t)$ is a piecewise linear function
of the form

\be
\omega^2(t)=\left\{ 
\begin{array}{l}
\omega_0^2 \quad \quad \quad \quad \quad   \quad{\rm if}\ \ t\le 0\\
 \omega_0^2+ \frac{(\omega_1^2-\omega_0^2)}T\, t \quad \  {\rm if}\ \  
0 <t < T\\
\omega_1^2 \quad \quad  \quad \quad \quad \quad  {\rm if}\ \ t\ge T
\end{array}. \right.
\ee
Thus $\omega(t)$ has discontinuous first derivative at $t=0$ and $t=T$,
and belongs to the class ${\cal C}^0$. Introducing the notation 
$\tilde  a=\omega^2_0, \ \tilde   b= \omega_1^2-\omega_0^2$ 
we obtain  that  on the interval $(0,T)$  the equation  (\ref{eq:Newton})
has the form 

\be \label{eq:eqlin}
\ddot q+(\tilde  a+\frac{\tilde  bt}T) q=0.
\ee
Two linear independent solutions of (\ref{eq:eqlin})
are given by the Airy functions:

\be
\psi_1(t)=
Ai(\frac{\tilde  bt+\tilde  a T}{\tilde b^{2/3}T^{1/3}})
\ee 
and

\be
\psi_2(t)=
Bi(\frac{\tilde  bt+\tilde  a T}{\tilde b^{2/3}T^{1/3}}).
\ee 
The  elements $a,b,c,d$ of the matrix $\Phi_T$ are defined by
the equation (\ref{eq:p1q1})  and (\ref{eq:phiT}).

The exact analytic expression for 

\be \label{eq:m1ap}
\mu^2 = 
\overline{(E_1-\bar E_1)^2}=\frac {E_0}2 \left(\delta^2+\gamma^2 \right)
\ee
is very complex, and we do not show it here. However, 
for $\omega_0^2=1,\omega_1^2=2$, $E_0=1$, using the asymptotic expansion
10.4.60,62,64,66 of \cite{Abramowitz} (pp.448-449), 
  we obtain the following  approximation 

\be \label{eq:m1ap1}
\overline{(E_1-\bar E_1)^2}
\approx 
\frac {\ep^2}{128}  \left( 9 - 4\,{\sqrt{2}}\,
\cos (\frac{4 - 8\,{\sqrt{2}}}{3\, {\ep}})  \right),
\ee
where we introduce {\bf the adiabatic parameter} $\ep$,

\be  \label{eq:adpar}
\ep=\frac 1T.
\ee
We see in figure 1 that the exact expression (\ref{eq:m1ap})
and its leading asymptotic approximation (\ref{eq:m1ap1}) practically
coincide, which demonstrates the power of the asymptotic expansion of the 
relevant expressions containing the Airy functions. Observe that the
decay of $\mu^2$ to zero as $\ep \rightarrow 0$ is oscillatory but
quadratic on the average, namely as $y=\frac 9{128}\epsilon^2$, which means
that $\mu$ goes to zero linearly with the adiabatic parameter $\ep$.
This is always the case when $\omega (t)$ is of class ${\cal C}^0$.
As we will see, in general, if $\omega (t)$ is of class ${\cal C}^m$,
then $\mu$ goes to zero oscillatory but in the mean as a power $\ep^{m+1}$.
This theorem will be proven in subsection 4.2 using the exact formulation
of the WKB method, applied to the relevant (but arbitrarily high) order.
We shall see in subsection 4.2, equation (\ref{eq:main},)  
that the leading WKB term 
precisely  reproduces the exact leading term in (\ref{eq:m1ap1}).
In figure 2 we show the same thing as in figure 1 but on the larger scale
of $\ep$.

\putfig{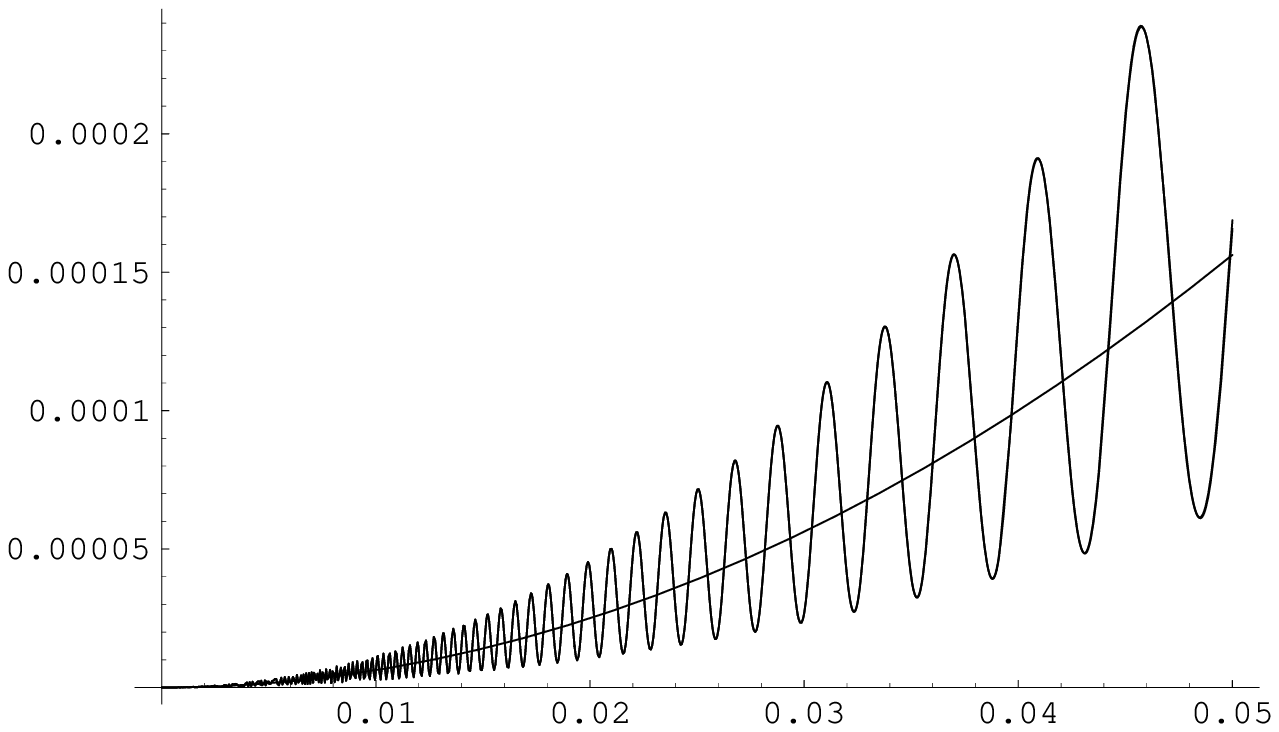}{3in}{5in}{$\overline{(E_1-\bar E_1)^2} $ for $ 0<\epsilon <0.05$; the lines of the exact expression (\ref{eq:m1ap}) 
and the asymptotics (\ref{eq:m1ap1}) practically coincide;
the non-oscillating thin line is the parabola $y=\frac 9{128}\epsilon^2$.
 }{hh}

\putfig{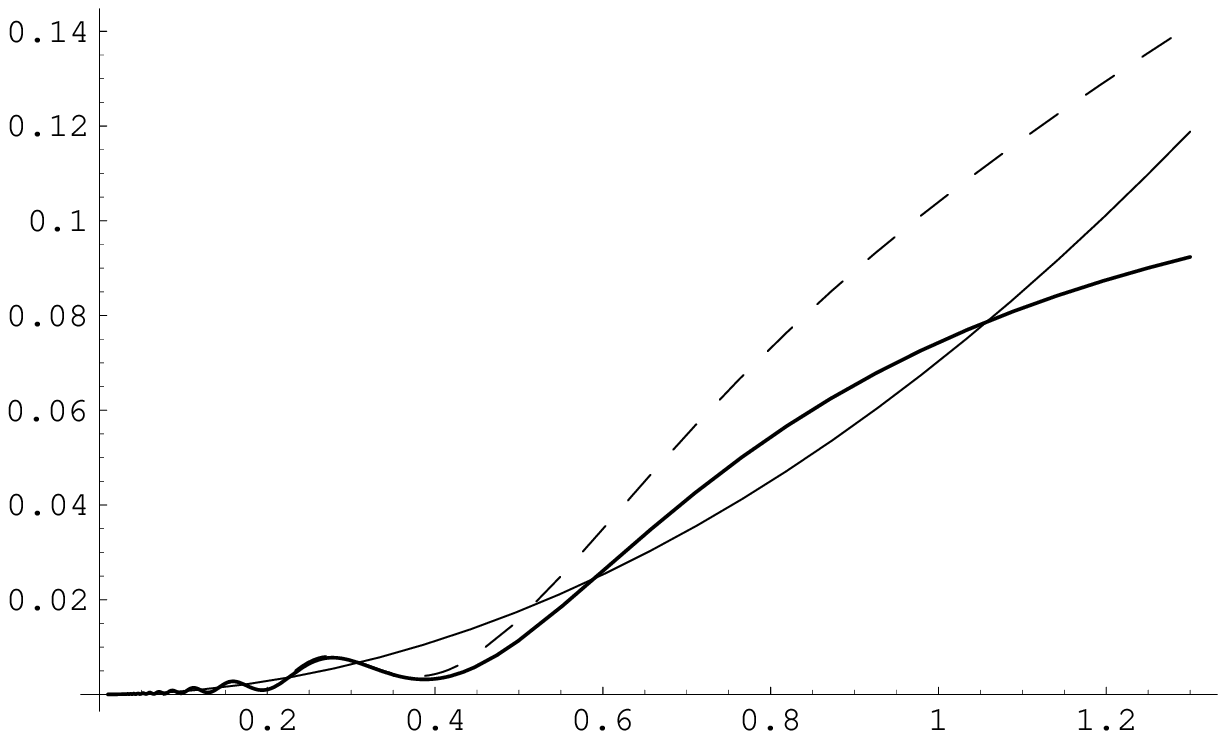}{3in}{5in}{$\overline{(E_1-\bar E_1)^2} $ for $ 0<\epsilon <1.2$;
the lines of the exact expression (\ref{eq:m1ap}) (full) 
and of the asymptotics (\ref{eq:m1ap1}) (dashed)  practically coincide 
for $\ep \le 0.3$;
the non-oscillating thin line is the parabola $y=\frac 9{128}\epsilon^2$.
One can show that $\mu^2$ goes to $1/8=0.125$ when $\ep \rightarrow \infty$,
which means $T\rightarrow 0$, which means the instantaneous jump of 
$\omega_0 =1$ to $\omega_1 = \sqrt{2}$.
 }{hh}

We now compute the final energy distribution function

$$
P(E_1)=
\frac 1{2\pi} \sum_{i=1}^4\left| \frac{d\phi}{dE_1} \right|_{\phi=\phi_j(E_1)}.
$$
Note that 

\be
\frac{d\phi}{dE_1} =  \frac{1}{E'_1(\phi)}.
\ee
Using the substitution
\be
u=\tan (\phi/2)
\ee
we obtain from (\ref{eq:ee1})

\be
(1+u^2)^2E_1=E_0 (\alpha (1-u^2)^2+4 \beta u^2+4 \gamma u (1-u^2)).
\ee
Resolving this equation with respect to $u$ yields four branches of the 
function $\phi(E_1)$. In figure 3 we show the emerging distribution function
as an illustration.

\putfig{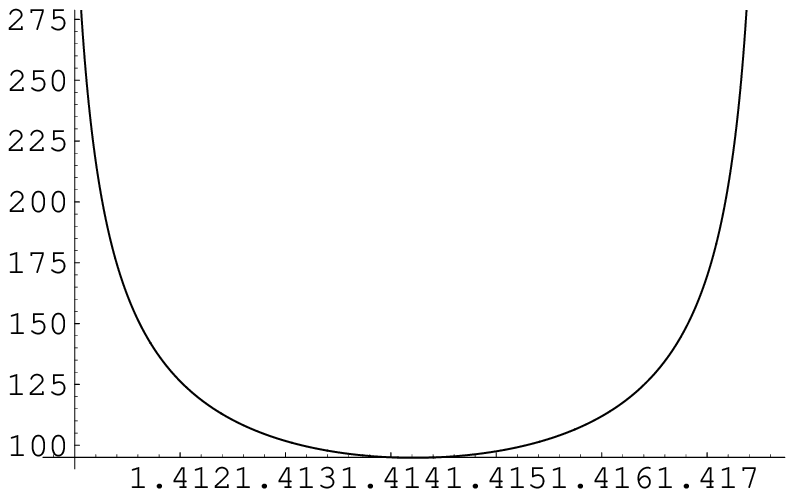}{3in}{5in}{The distribution function $P(E_1)$ for
the linear model of subsection 3.1, for $E_0=1, \omega_0^2=1,
\omega_1^2=2, \epsilon=0.01$.
 }{hh}

\subsection{The harmonic model: class ${\cal C}^1$}

Now we consider the case when $\omega (t)$ is of class ${\cal C}^1$, namely
we use the specific model

\be
\omega^2(t)=\left\{ 
\begin{array}{l}
\tilde  a \quad \quad \quad \quad \quad \quad \quad \quad {\rm if}\ \   t\le 0\\
\tilde b -  (\tilde   b -\tilde a)\,\cos (\frac{\pi \,t}{T})\
  0 <t < T\\
2\tilde  b - \tilde a   \quad \quad \quad \quad \quad \quad {\rm if}\ \ t\ge T
\end{array}, \right.
\ee
where  $\tilde a=\omega^2_0, \ 2\tilde  b - \tilde a = \omega_1^2$.
Then the equation  (\ref{eq:Newton}) has the form  

\be \label{eqcos}
\ddot q+ \left( \tilde b-(\tilde b-\tilde a)\,\cos\left( \frac{\pi t}{T}
\right)\right) q=0
\ee
and the fundamental solutions are represented by 
the Mathieu functions:

\be
\psi_1(t)=ce(\frac{4 \tilde b T^2}{{\pi }^2},\frac{2 (\tilde b- \tilde
a) T^2}{{\pi }^2},
\frac{\pi t}{2 T})
\ee 
and
\be
\psi_2(t)=se(\frac{4 \tilde b T^2}{{\pi }^2},\frac{2 (\tilde b-\tilde a) T^2}
{{\pi }^2}, \frac{\pi \,t}{2 T}).
\ee 
In this model the first derivative of $\omega (t)$ vanishes at
$t=0$ and $t=T$ and is continuous there, whilst the second derivative is
discontinuous, therefore $\omega (t)$ belongs to the class ${\cal C}^1$.

Here we find that indeed $\mu^2$ goes to zero oscillating but in
the mean as $\ep^4$, as shown in figure 4. The exact analytic asymptotic
behaviour of the Mathieu functions needed for our calculations is
not known, unlike for the Airy functions, 
but we compare the exact (numerical result) with the
WKB approximation expounded in section 4, see the final general formula
(\ref{eq:mainfinal}) of subsection 4.3.

\putfig{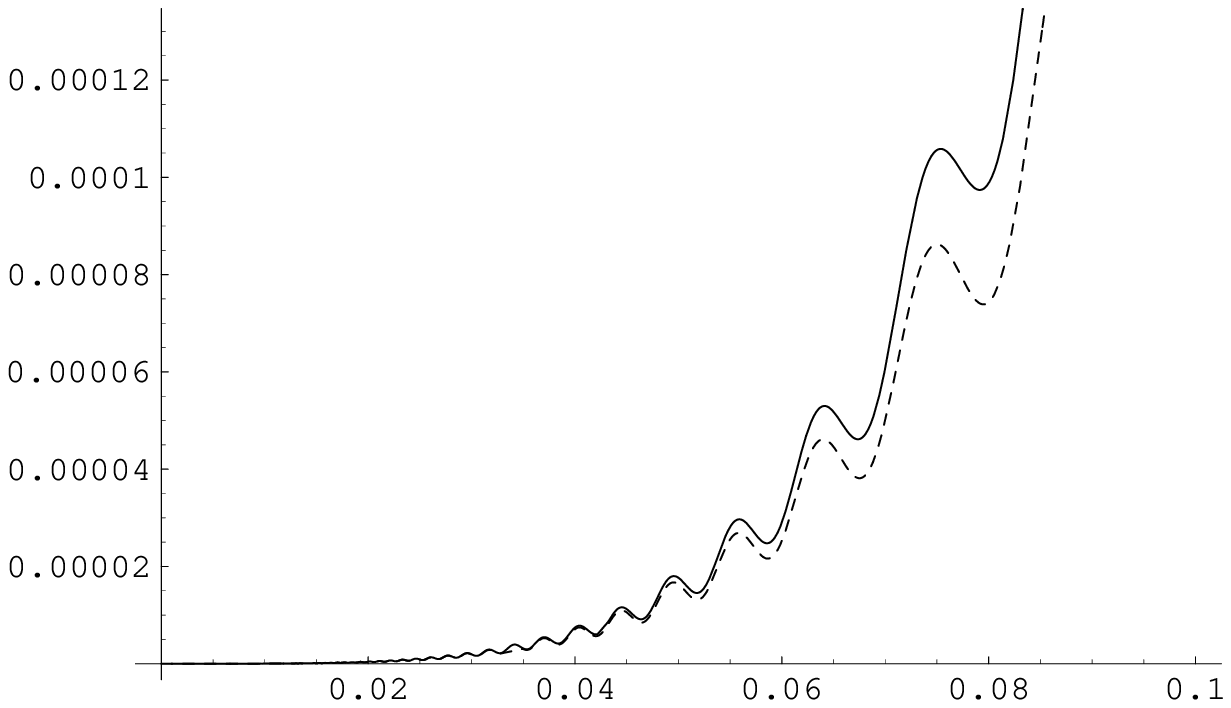}{3in}{5in}{We show, for the harmonic model of 
subsection  3.2,
$\overline{(E_1-\bar E_1)^2} $ for $ 0<\ep <0.1$.
The exact result is represented by the full line, whilst
the dashed curve is the curve  
$0.056\, {\ep}^4\,\left( 41 + 9 \,\cos (\frac{2.78}{\ep}) \right)$,
obtained by the WKB method, equation (\ref{eq:mainfinal}) of
subsection 4.3. }{hh}

\subsection{The analytic model: class ${\cal C}^{\infty}$}

Consider now the case when

\be \label{eq:expmodel}
\omega^2(t)=
\frac{1 + a\,e^{{\alpha}t}}{1 + e^{{\alpha}t}},
\ee
that is,  ( \ref{eq:Newton}) is of  the form 

\be \label{eq:exp}
\ddot q +
\frac{1 + a\,e^{{\alpha}t}}{1 + e^{{\alpha}t}}  q=0.
\ee
(Please do not confuse these $a$ and $\alpha$ with the usage in the main text.)
Then two fundamental solutions of (\ref{eq:exp})
  are represented using the hypergeometric function $_2F_1$ as follows:

\be\label{eq:ef1}
\psi_1(t)=e^{-\imag t}\, { {_2F_1}(-\frac{\imag }{\alpha} - \frac{\imag 
\,{\sqrt{a}}}{\alpha},-
     \frac{\imag }{\alpha} + \frac{\imag \,{\sqrt{a}}}{\alpha},1 - 
\frac{2\,\imag }{\alpha},-e^{\alpha\,t})}
\ee
and

\be\label{eq:ef2}
\psi_2(t)= e^{\imag  t }\,
    {_2F_1}(\frac{\imag }{\alpha} - \frac{\imag \,{\sqrt{a}}}{\alpha},
    \frac{\imag }{\alpha} + \frac{\imag \,{\sqrt{a}}}{\alpha},1 + 
\frac{2\,\imag }{\alpha},-e^{\alpha\,t}).
\ee

\putfig{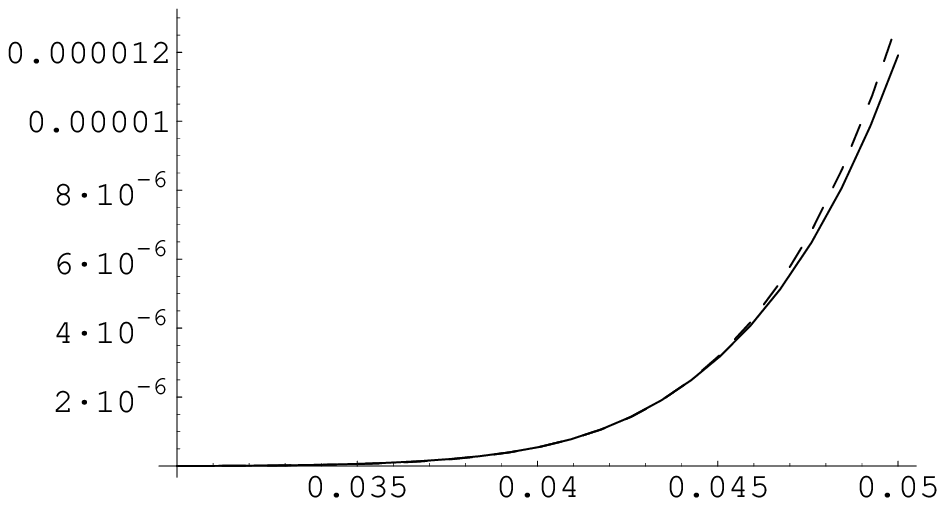}{3in}{5in}{
The variance $\overline{(E_1-\bar E_1)^2} $ of the energy 
for the analytic model (\ref{eq:expmodel}),  for $ 0.03<\ep<0.05$.
The dashed curve   is approximation  $y=4.174 e^{-0.634/\ep}.$  }{hh}
\noindent
Using (\ref{eq:ef1}) and (\ref{eq:ef2})   we compute  the  matrix $\Phi$ 
defined by (\ref{eq:phiT})
with $t_0=-\frac{10}\alpha $ and   $t_1=\frac{10}\alpha $,
so $T=t_1-t_0=\frac {20}\alpha$ (note that the most essential change of the
function $\om(t)$ is on the interval $(-\frac2 \alpha,\frac 2  \alpha)$).
For the case $a=2$ the graph of {$\overline{(E_1-\bar E_1)^2}$ as the 
function
of $\ep=1/T$ is displayed in figure 5  as continuous line.
The approximation to the graph,  $y=4.174 e^{-0.634/\ep}$, computed 
using the least square method,
is displayed in the figure as  dashed line. We clearly see that
the behaviour of $\mu^2$ is exponential.

\section{Application of the WKB method for the calculation of 
the transition map}

\subsection{General and exact considerations}

In this section we 
 proceed with the calculation of the transition map $\Phi$ in
the general case, and because (\ref{eq:Newton})
is generally not solvable,
we have ultimately to resort to some approximations. Since the adiabatic
limit $\ep \rightarrow 0$ is the asymptotic regime that we would
like to understand, the application of the rigorous WKB theory
(up to all orders) is most convenient, 
and usually it turns out that the leading
asymptotic terms are well described by just the leading WKB terms
if $\ep$ is sufficiently small. In using the WKB method we 
refer to our work \cite{RR2000}, where we have derived 
the explicit analytic expressions for all WKB orders in closed form,
except for the exact rational coefficients, which can be easily
obtained from a recurrence formula.
\footnote{There is substantial literature on WKB method, which
due to limited space cannot be reviewed here. But we should mention the
classic works by N. Fr\"oman and  P.O. Fr\"oman,  who have found a number
of interesting relationships, e.g. a relation between the even and odd 
order terms \cite{Froman}, although we do not use it here, 
so that our exposition is selfcontained.}

We introduce re-scaled and dimensionless time  $\lambda$
\be
\lambda= \epsilon t,  \;\;\;\;\; \ep = 1/T,
\ee
so that (\ref{eq:Newton}) is transformed to the equation


\be \label{eq:sch}
\ep^2 q''(\la)+\omega^2(\lambda)q(\la)=0.
\ee
By prime we denote the differentiation w.r.t. $\lambda$.
Let $q_+(\lambda)$ and $q_-(\lambda)$  
be two linearly  independent solutions of 
(\ref{eq:sch}). Then the matrix  (\ref{eq:phig}) takes the form

\be \label{eq:phig1}
\Psi_\la=\left(
\ba{cc}
 q_+(\la) & q_-(\la)\\
\ep M q'_+(\la) & \ep M  q'_-(\la)
\ea
\right)
\ee
and taking into account that $\la_0=\ep t_0, \la_1=\ep t_1 $ we obtain for the 
matrix (\ref{eq:transitionmap}) the expression

\be 
\Phi=\left(
\ba{cc}
 a & b\\
  c & d
\ea
\right)= \Psi_\la(\la_1)  \Psi_\la^{-1}(\la_0).
\ee
We now use the WKB method in order to obtain 
the coefficients $a, b, c, d$ of the matrix $\Phi$.
To do so,  we look for solution of (\ref{eq:sch}) in the form

\be
q(\la) = w \exp \left\{\frac 1\epsilon \sigma (\la)\right\}
\ee
where  $\sigma (\la)$ is a complex function that satisfies the
differential equation 

\be \label{eq:w1}
(\sigma '(\la))^2+ \epsilon  \sigma '' (\la)=-\om^2(\la)
\ee
and $w$ is some constant with dimension of length.
The WKB expansion for the phase is

\be \label{eq:w2}
\sigma(\la)=\sum_{k=0}^{\infty}
  \epsilon^k \sigma_k (\la).
\ee
Substituting (\ref{eq:w2}) into (\ref{eq:w1}) and comparing
like powers of  $\epsilon$ gives the recursion relation

\be \label{eq:w3}
\s_0'^2 = -\omega^2(\la),\ \ \ \ \ \s'_{n}=-\frac 1{2\s'_0}
(  \sum_{k=1}^{n-1} \s' _k \s ' _{n-k}+\s_{n-1}'').
\ee
Here we apply our WKB notation and formalism from our work
\cite{RR2000} and we can choose
$\s'_{0,+} (\la)=\ii \om(\la)\quad {\rm  or} \quad  \s'_{0,-} (\la)
=-\ii \om(\la)$.
That  results in two linearly independent solutions of (\ref{eq:sch})  
given by the WKB  expansions with the coefficients

\bea
\s_{0,\pm}(\la)=\pm \ii  \int _{{\lambda_0}}^{\la} \om (x) dx, 
\;\;\; \s_{1,\pm}(\la)=
- \frac 12 \log \frac{\om(\la)}{\om(\la_0)}, \\ \s_{2,\pm}= \pm
\frac{\imag }{8}\,\int _{{\lambda_0}}^{\la}
    \frac{3\,{\om'(x)}^2 - 2\,\om(x)\,\om''(x)}
      {{\om(x)}^3}\,dx, \ \dots
\eea
Since $\om (\la)$ is a real function we deduce from (\ref{eq:w3}) 
that all functions $\s'_{2k+1}$ are real and all   functions $\s'_{2k}$
are pure imaginary
and 
$\s'_{2k,+}=- \s'_{2k,-}, \quad \s'_{2k+1,+}= \s'_{2k+1,-}$
where $k=0,1,2,\dots$, and thus we have 
$\s'_+=A(\la)+\ii B(\la), \quad \s'_-=A(\la)-\ii B(\la)$
where 
$A(\la) = \sum_{k=0}^\infty \ep^{2k+1}\s'_{2k+1}(\la), \quad 
B(\la) = -\ii \sum_{k=0}^\infty \ep^{2k}\s'_{2k,+}(\la)$
are both real quantities.
 Integration of the above equations 
yields

\be
\s_+=r(\la)+\ii s(\la), \quad \s_-=r(\la)-\ii s(\la),
\ee
where 
$r(\la)=\int_{\la_0}^\la A(x)\, dx, \quad  s(\la)=\int_{\la_0}^\la B(x)\, dx$.
Below we shall denote $s_1=s(\la_1)$. 
To simplify the expressions  let us denote  $A_0= A(\la_0)$,
$A_1= A(\la_1)$, $B_0= B(\la_0)$ and $B_1= B(\la_1)$.

Using this notation we find that the elements of the matrix $\Phi_\la$ 
have the following form:

\bea \label{eq:abcd}
a=  &- \frac{e^{\frac {r(\lambda_1)}\ep}}{\bn} \left[\an \si -\bn \co \right],
\\ \nonumber
b=& \frac{\ek}{M \bn}\si,\\ \nonumber
c=& - \frac M\bn \ek \left[ \left (\an \ak +\bn \bk \right)\si \right. 
\\ \nonumber
& \left. 
+ \left( \an \bk - \ak \bn   \right)\co  \right],\\ \nonumber
d=&  \frac \ek\bn \left[ \ak \si +\bk \co  \right]. 
\eea
Therefore for the quantities $\alpha$ and $\beta$ defined by (\ref{eq:abg})
we obtain

\bea
\alpha=&   \frac \ekd{\on^2 \bn^2} \left[ \ok^2 \left( \an \si -\bn 
\co  \right)^2 + \right.\\  \nonumber
& 
 \left(  \left( \an \ak + \bn \bk  \right)\si + \right.\\ \nonumber
& \left. \left.
 \left( \an \bk - \ak \bn  \right)\co   \right)^2   \right],
\eea

\bea
\beta= \frac{\ekd}{\bn^2} \left[ \ok^2 (\si )^2 +\left( \ak \si +\bk \co  
\right)^2  \right]. 
\eea
We remind that $\det (\Psi)=1$.  Computing now   $\det (\Psi)=ad-bc$
using the expressions (\ref{eq:abcd})
we find

\be
\left(
\frac { \ek}{  \bn}\right)^2=\frac 1{\bn \bk}
\ee
and therefore finally, exact to all orders,

\bea \label{eq:abcdf}
a = & -\frac{1}{\sqrt{\bn\bk}} \left[\an \si -\bn \co \right],\\ \nonumber
b = &  \frac{1}{M \sqrt{\bn\bk}}\si,\\ \nonumber
c = & - \frac{M}{\sqrt{\bn\bk}  }\left[ \left (\an \ak +\bn \bk \right)\si \right. \\ \nonumber
& \left. 
+ \left( \an \bk - \ak \bn   \right)\co  \right],\\ \nonumber
d = &  \frac{1}{\sqrt{\bn\bk}  }\left[ \ak \si +\bk \co  \right]. 
\eea
Thus we obtain the final result for the expression (\ref{eq:bare1}), 
exact to all orders, 

\bea \nonumber
\alpha+\beta  =  \frac 1{\bn \bk}
\left[
 \siq \left( \frac{\bn^2 \bk^2}{\on^2}+\ok^2   \right)
+\coq \left( \bn^2 \frac{\ok^2}{\on^2}+\bk^2    \right)
+  \right.  \\ \label{eq:apb}
\siq  
 \left( \an^2 \frac{\ok^2}{\on^2}+\frac{\an^2 \ak^2}{\on^2}+
\frac {2 \an \ak \bn \bk}{\on^2}+\ak^2  \right)+ 
\\  \nonumber
\coq \left(\frac {\an^2\bk^2}{\on^2}+\frac {\ak^2\bn^2}{\on^2}- 
  \frac {2 \an \ak \bn \bk}{\on^2} \right)
+ \\ \nonumber
 \si \co\times  \\ \nonumber
  \left. \left(  -2 \an \bn \frac{\ok^2}{\on^2}+ 2 \ak \bk 
 +\frac 2{\on^2} \left(
\an \ak +\bn \bk \right) \left(\an \bk -\ak \bn  \right) \right) 
\right].
\eea

\subsection{Leading asymptotic terms in the power expansion in terms of $\ep$}

So far the result is exact. Let us consider the first order WKB approximation,
which is the generic case, that is

\be
A(\la)\approx \ep \s'_{1,+}(\la), 
\quad B (\la) \approx \frac{\s'_{0,+}(\la)}\ii=\om(\la). 
\label{eq:ABfirst}
\ee  
Substituting these values of $A(\la) $ and $B(\la) $ into (\ref{eq:apb})
we find

\bea \label{eq:apb1}
\alpha+\beta=2 \frac \ok\on + \\ \nonumber
\ep^2 \left( \frac {\ok \onp^2 }{4 \on^5}-
\frac{\cos\left( \frac {2\int_{\la_0}^{\la_1} \om(x) \,dx}\ep \right)\onp 
\okp }{2 \on^3 \ok } + \frac{\okp^2}{4 \on \ok^3}  \right)+O(\ep^3).
\eea
As we have shown above $\bar E_1=\frac {\alpha+\beta}2 E_0$
and $ \frac {\overline{(\Delta E_1)^2}}{E_0^2}= \frac 12\left[ 
\left(  \frac{\bar E_1}{E_0}  \right)^2 - 
\left( \frac \ok\on \right)^2   \right] $. Therefore

\bea \label{eq:main}
 \frac {\overline{(\Delta E_1)^2}}{E_0^2}= \\ \nonumber
\ep^2 \left(
 \frac {\ok^2 \onp^2 }{8 \on^6}-
\frac{\cos\left( \frac {2\int_{\la_0}^{\la_1} 
\om(x) \,dx}\ep \right)\onp \okp }{4  \on^4  } + 
\frac{\okp^2}{8 \on^2 \ok^2}  \right)+O(\ep^3). 
\eea
Substituting into the last formula $\om(\la)= \sqrt{1+\la} $
we obtain exactly the approximation (\ref{eq:m1ap1}). Thus the WKB 
approach yields exactly the leading asymptotic term for $\bar E_1$
and $\mu^2$ for general $\omega (t)$, and not only that, but also
all the higher power terms of $\ep$ if desired. Below we shall
explain how all the higher order WKB terms can be calculated using
our  closed form  formula \cite{RR2000}. Please note that
the expression in the big brackets of (\ref{eq:main}) is
positive definite (its minimal value is a complete square of a real
quantity), as it must be.

Now we consider the special cases where $\onp$ and $\okp$ vanish which
implies that the leading order terms in the above two equations 
vanish. In such a special case we have to work out the higher order
WKB terms. By this reasoning and calculation we prove that if $\omega (t)$ 
is of class ${\cal C}^{m}$ (all derivatives up to and
including the order $m$ are continuous) then $\mu \propto \ep^{(m+1)}$.

Therefore we now suppose that all derivatives at $\la_0$ and $\la_1$
vanish up to order $(n-1)$

\be
\om'(\la_0)=\dots = \om^{(n-1)}(\la_0)=\om^{(n-1)}(\la_1)=0,\
\om^{(n)}(\la_0)\om^{(n)}(\la_1)\ne 0.
\ee 
Then
\be
\s'_1(\la_0)=\s'_1(\la_1)=\dots = \s'_{n-1}(\la_0)=\s'_{n-1}(\la_1)=0,\
\s'_{n}(\la_0)\s_{n}(\la_1)\ne 0.
\ee 
There are two cases to be considered separately, namely,
the case of odd and the case of even  $n$.

In the case $n=2k-1$ in order to find an approximation 
to $\mu^2$ up to order $\ep^{4k-2} $  we expand $A(\la)$ 
and $B(\la)$ as follows:

\bea \label{eq:ABappr}
A(\la) & = & \sum_{u=k}^{2k}  \ep^{2u -1} \s'_{2u-1,+}(\la)+O(\ep^{4k+1}), \\
 \nonumber
B (\la) & = & \om(\la)-\ii \sum_{u=k}^{2k-1} 
\ep^{2u}  { \s'_{2u,+}(\la)} +O(\ep^{4k}) .
\eea
We observe that the terms of the approximation (\ref{eq:ABappr})
of the order higher than $\ep^{2k}$ are essential  only in the following 
 terms appearing in the brackets of the expression (\ref{eq:apb}):

\bea \label{eq:terms} 
\frac{\coq \ok^2 \bn^2}{\on^2},\quad \coq \bk^2,\quad  
\frac{\siq \bn^2 \bk^2}{\on^2},\\ \nonumber
\si\co \left(-2 \frac{ \ok^2 \an \bn }{\on^2}+ 
2 \ak  \bk -2  \frac{ \ak \bk  \bn^2   }{\on^2  }+
 2  \frac{ \an \bn  \bk^2   }{\on^2  } \right) .
\eea
Taking into account that

\be
B(\la)^2=\om(\la)^2 -2 \om(\la)  \ii 
\sum_{u=k}^{2k-1} \ep^{2u}  { \s'_{2u,+}(\la)}+O(\ep^{4k}),   
\ee
\be 
B_0^2 B_1^2 =\on^2 B_1^2+\ok^2 B_0^2 -\on^2 \ok^2 +O(\ep^{4k}),
\ee
and

\be
A(\la) B(\la) =\om(\la) A(\la) +O(\ep^{4k-1})
\ee
we easily find:

\be 
-2 \frac{ \ok^2 \an \bn }{\on^2}+ 2  \frac{ \an \bn  \bk^2   }{\on^2  }=
O(\ep^{4k-1}),
\ee
\be
 2 \ak  \bk -2  \frac{ \ak \bk  \bn^2   }{\on^2  }=O(\ep^{4k-1}),
\ee
yielding that the coefficient of $\si \co $ in 
(\ref{eq:terms}) is of the order $O(\ep^{4k-1})$, and we find

\bea
\frac{\coq \ok^2 \bn^2}{\on^2}+ \coq  \bk^2+  \frac{\siq \bn^2 \bk^2}{\on^2}=
\\ \nonumber
2 \ok^2 \coq +\ok^2 \siq+\\
\nonumber 
\sum_{u=k}^{2k-1} \ep^{2u} \left( - 
2 \ok \ii  { \s'_{2u,+}(\la_1)}- 2 \ii \frac{\ok^2}\on  \s'_{2u,+}
(\la_0)\right)+ O(\ep^{4k-1}).
\eea
Therefore the expression in the brackets of (\ref{eq:apb}) is equal to

\bea  \label{eq:abpr}
  2 \ok^2 + \sum_{u=k}^{2k-1} \ep^{2u} \left( - 2 \ok \ii  { \s'_{2u,+}(\la_1)}- 2 \ii \frac{\ok^2}\on  \s'_{2u,+}(\la_0)\right) +\\ \nonumber
\ep^{4k-2}\left(
\frac {\ok^2   \s'_{2k-1,+}(\la_0) ^2}{ \on^2}
+{\  \s'_{2k-1,+}(\la_0) ^2} -   \right. \\ \nonumber  \left.  \frac{2 \ok  \s'_{2k-1,+}(\la_0)  \s'_{2k-1,+}(\la_1) }{\on} \left(1-2 \siq \right)\right)
 +O(\ep^{4k-1}). 
\eea
Noting that 

\be
 \frac 1{\bn \bk}=\frac 1{\on \ok} \left(1+ \ii \sum_{u=k}^{2k-1} 
\ep^{2 u} \left( 
 \frac {\s'_{2u,+}(\la_0)}
\on +\frac {\s'_{2u,+}(\la_1)}\ok
 \right) \right) +O(\ep^{4k})
\ee
we finally obtain 

\bea \label{eq:sim1}
  \frac {\overline{(\Delta E_1)^2}}{E_0^2}=  
\ep^{4k-2}\left(
\frac {  \s'_{2k-1,+}(\la_1) ^2}{2 \on^2}+\frac{\ok^2  
\s'_{2k-1,+}(\la_0) ^2}{2\on^4} - \right. \\ \nonumber
\left.  \frac{\ok  \s'_{2k-1,+}(\la_0)  \s'_{2k-1,+}(\la_1) }{\on^3}
\cos\left( \frac {2 s_1 }\ep\right)\right)+O(\ep^{4k-1}). 
\eea

The second case is when $n$ is even, $n=2k$, $k$ positive integer,
so that we assume now

\bea \label{eq:ABeven}
A(\la)= \sum_{u=k}^{2k} \ep^{2u+1} \s'_{2u+1,+}(\la) + h.o.t., \\ \nonumber
B (\la) = \om(\la)-\ii \sum_{u=k}^{2k} \ep^{2u} \s'_{2u,+}(\la) + h.o.t.
\eea
Then, similarly as above, using the equalities

\be
B(\la)^2=\om(\la)^2 -2 \om(\la)  \ii \sum_{u=k}^{2k-1} \ep^{2u}  
{ \s'_{2u,+}(\la)} -\ep^{4 k} (\s'_{2k,+}(\la) )^2 +O(\ep^{4k+2}),   
\ee
\be 
B_0^2 B_1^2 =\on^2 B_1^2+\ok^2 B_0^2 -\on^2 \ok^2 -4 \on \ok 
\ep^{4 k} \s'_{2k,+}(\la_0)  \s'_{2k,+}(\la_1)  +O(\ep^{4k+2}),
\ee
\be
A(\la) B(\la) =\om(\la) A(\la) +O(\ep^{4k+1})
\ee
and

\bea
 \frac 1{\bn \bk}=\frac 1{\on \ok} \left(1+ \ii \sum_{u=k}^{2k} 
\ep^{2 u} \left( 
 \frac {\s'_{2u,+}(\la_0)}
\on +\frac {\s'_{2u,+}(\la_1)}\ok
 \right) \right) - \\   \nonumber
\ep^{4 k}\left(
 \frac {(\s'_{2k,+}(\la_0))^2}{\on^2} +  \frac {(\s'_{2k,+}(\la_0))
 (\s'_{2k,+}(\la_1))} {\on \ok  }+
 \frac {(\s'_{2k,+}(\la_1))^2}{\ok^2} \right)  +O(\ep^{4k+1})
\eea
we obtain

\bea \label{eq:sim2}
  \frac {\overline{(\Delta E_1)^2}}{E_0^2}=  -
\ep^{4k}\left(
\frac {  \s'_{2k,+}(\la_1) ^2}{2 \on^2}+\frac{\ok^2  \s'_{2k,+}
(\la_0) ^2}{2\on^4}- \right. \\ \nonumber
\left.  \frac{\ok  \s'_{2k,+}(\la_0)  
\s'_{2k,+}(\la_1) }{\on^3}\cos\left( \frac {2 s_1 }\ep\right)\right)+
O(\ep^{4k+1}). 
\eea
We now observe that the expression 

\bea \label{eq:simm}
  \frac {\overline{(\Delta E_1)^2}}{E_0^2}= ( -1)^{n+1}
\ep^{2n}\left(
\frac {  \s'_{n,+}(\la_1) ^2}{2 \on^2}+\frac{\ok^2  
\s'_{n,+}(\la_0) ^2}{2\on^4}- \right. \\ \nonumber
\left.  \frac{\ok  \s'_{n,+}
(\la_0)  \s'_{n,+}(\la_1) }{\on^3}\cos\left( \frac {2 s_1 }\ep\right)
\right)+O(\ep^{2n+1}). 
\eea
coincides with  (\ref{eq:sim1}) when $n=2k-1$ and with (\ref{eq:sim2}) when 
$n=2k$. If $n$-th derivative of $\omega (t)$ at $\la_0$ and $\la_1$
is nonzero, it means that $\om (t)$ is of class ${\cal C}^m$ where
$m=n-1$, or $n=m+1$. Then $\mu^2$ is oscillating as 
$\ep \rightarrow 0$ but in the mean goes to zero as a power
$\mu^2 \propto  \ep^{2(m+1)}$, and therefore 
the general equation (\ref{eq:simm}) proves our assertion.

\subsection{Further simplifications of the general formula for the leading
asymptotic term (\ref{eq:simm})}

Now we want to simplify the expression (\ref{eq:simm}) by expressing
$\s'_n$ in terms of $\om'_n$, using our explicit results in \cite{RR2000}.

Let $M=\cup_{k=1}^\infty{\bf N}^k$, ${\bf N}$ is the set of non-negative integers.
We define the  map $L : M \to {\bf N}$ by 

\be \label{eq:lk}
L(\nu)= 1\cdot \nu_1+2\cdot \nu_2 +\dots + l \cdot \nu_l
\ee
and denote by $L(\nu)=m$ the equation 

\be \label{eq:eqm}
L(\nu)= 1\cdot \nu_1+2\cdot \nu_2 +\dots + m \cdot \nu_m=m,
\ee
with $m\in {\bf N}$, $\nu\in M$ (there is one-to-one correspondence between the set $\{(\nu_1,\dots,\nu_m)\}$ of solutions of (\ref{eq:eqm}) and the set of  \emph{partitions}
of $m$).
 For a vector $\nu=(\nu_1,\dots, \nu_l)\in M $ we denote   $Q^{(\nu)}=
(Q')^{\nu_1}(Q'')^{\nu_2}\dots (Q^{(l)})^{\nu_l}$, $|\nu|=\nu_1+\dots+\nu_l$
 and let  $\nu(i)$ $ (i=1,\dots,l-1)$ be 
the vector
$(\nu_1, \dots, \nu_{i}+1,\nu_{i+1}-1,\dots,\nu_{l}).$
It is  shown in \cite{RR2000}  that the functions $\s'_m$  are of the form

\be \label{eq:sk}
\s'_m=\sum_{\nu: L(\nu)=m} \frac{U_\nu
Q^{m-|\nu|}Q^{(\nu)}}{Q^{\frac{3m-1}{2}}},
\ee
where   the coefficients $U_\nu $ satisfy the
recurrence relation

\bea \label{eq:u}
U_\nu=  \\ \nonumber
\frac 12 \sum_{\mu,\theta\ne 0, \mu+\theta=\nu} U_\mu U_\theta +
\frac{(4-L(\nu)-2|\nu|)U_{(\nu_1-1,\nu_2,\dots,\nu_{l})}}4+\sum_{i=1}^{l-1}
\frac{(\nu_i+1)U_{\nu(i)}}2,
\eea
with  $U_{\bar 0}=-1$ and we put $U_\alpha=0$ if among the coordinates of 
the vector $\alpha$ there is a negative one.

In our case $Q(x)=-\omega^2(x)$. Therefore
in the case when 
\be \label{eq:omp0}
\om'(\la)=\dots = \om^{(n-1)}(\la)=0,\
\om^{(n)}(\la)\ne 0
\ee 
from (\ref{eq:sk}) we obtain 

\be
\s'_n=\frac{U_{\tilde \nu}Q^{(n)}}{Q^{\frac{n+1}{2}}},
\ee
where $U_{\tilde \nu}$ is computed by (\ref{eq:u}) and $\tilde \nu=(0,\dots,0,1)$ is the solution of $L(\nu)=n$ with all entries except of the last one equal to zero. 
From  (\ref{eq:u}) we obtain 
$
U_{\tilde \nu}= \frac 1{2^{n-1}} U_{(1,0,\dots,0)}=\frac 1{2^{n-1}} \left( \frac 14+\frac 12 U_{\bar 0}  \right)=- \frac 1{2^{n+1}}
$
yieding
\be \label{eq:ssn}
\s'_n=- \frac{Q^{(n)}}{2^{n+1}Q^{\frac{n+1}{2}}}.
\ee
If (\ref{eq:omp0}) holds then 

\be
Q^{(n)}=- 2 \om \om^{(n)}.
\ee
Therefore, from (\ref{eq:ssn}) we obtain

\be
\s'_{n,+} = \frac{\ii^{n+1} \om^{(n)}}{2^{n}\om^{n}}.
\ee
Substituting this expression in  (\ref{eq:simm})  we obtain
the final general formula for the leading asymptotic term of
$\mu^2$, namely

\bea
 \label{eq:mainfinal}
\frac {\overline{(\Delta E_1)^2}}{E_0^2} = \\ \nonumber
\frac{\ep^{2n}}{2^{2n+1}} \left(
\frac{\ok^2 (\om_0^{(n)})^2 }{\on^{2(n+2)}}
+   \frac{ (\om_1^{(n)})^2 }{(\ok)^{2n}\on^2 }
 - 2  \frac{ \om_0^{(n)}\om_1^{(n)} }
{\on^{n+3}\ok^{n-1} } \cos\left( \frac {2 s_1}\ep\right) \right)+O(\ep^{2n+1}).
\eea
In the special case $n=1$ we recover the formula (\ref{eq:main}).
We emphasize that the above expression (\ref{eq:mainfinal}) is 
indeed evidently positive definite, which must always be the case
for the variance. 

If the lowest nonvanishing derivative is the 
$n$-th derivative of $\omega (t)$ at $\la_0$ and/or $\la_1$,
then it means that $\om (t)$ is of class ${\cal C}^m$ where
$m=n-1$, or $n=m+1$. Then $\mu^2$ is oscillating as 
$\ep \rightarrow 0$ but in the mean goes to zero as a power
$\mu^2 \propto  \ep^{2(m+1)}$, and therefore 
the general equation (\ref{eq:mainfinal}) proves our assertion.

If $\omega (t)$ is an analytic function on the real time
axis $(-\infty,+\infty)$, 
the decay to zero is oscillating and on the average is
exponential $\propto \exp (-const/\ep)$ \cite{LL}, \cite{Knorr},
\cite{Meyer1}, \cite{Meyer2}. This exponential smallness
stems from the divergence of the relevant series and has been
extensively studied in related works \cite{Berry1982},
\cite{Berry1990}, \cite{Joye1993}, \cite{Joye1991} and \cite{Lim1991},
where the resummation techniques have been devised.


Let us now summarize our results for the variance $\mu^2$ 
as a function of $\omega (t)$ embodied in the exact
general formulae (\ref{eq:abcdf}) and (\ref{eq:apb}). 
If $\omega (t)$ is analytic
between $t_0$ and $t_1 \ge t_0$ then the equation (\ref{eq:mainfinal})
applies, and as we see $\mu^2$ is dominated by the switch-on
and switch-off events at $t_0$ and $t_1$, respectively.
However, the smaller the lowest nonvanishing derivatives of $\omega (t)$ 
at the two points, the smaller will be the power law contribution.
Indeed, if $t_0$ and $t_1$ go to $-\infty$ and $+\infty$, respectively,
and $\omega (t)$ is analytic on the entire interval, then
the behaviour will become exponential at sufficiently large 
$\epsilon \ge \epsilon_c$,
whilst it will be oscillatory and in the mean a power law
at small $\epsilon \le \epsilon_c$. For example, in figure 
5 we observe only exponential regime, because $\epsilon_c$ is
so small.

If there is any non-analyticity of $\omega (t)$ on this interval, 
say at $t_i$, then  the calculation of $\mu^2$ must be
split into two intervals $(t_0, t_i)$ and $(t_i, t_1)$,
calculating the transition matrices for each interval
separately, then looking at their product and calculating
its matrix elements $a,b,c,d$, and then proceeding to
calculate $\alpha + \beta$ and finally
$\mu^2$, which  will obey a power law like in (\ref{eq:mainfinal}).

In other words, if $\omega (t)$ is analytic everywhere, and if the
switch-on and switch-off events are completely eliminated,
in the sense that they are infinitely smooth,
the behaviour of $\mu^2 (\epsilon)$ is exponential, as explained above.
In all other cases it is a power law.

\section{Periodic $\omega (t)$}

In case when $\om (t)$ is periodic with period $\tau$ but otherwise
completely general we can state some general rigorous results. 
Since  the frequency $\om_0$ at time $t_0$ and $\om_1$ at time
$t_1=t_0 + \tau$ are equal, 
we see from equation (\ref{eq:sigma}) due to positive definiteness
of $\mu^2$ that $\bar{E_1}$ is always greater than $E_0$, that is,
in a period $\tau$, or any integer multiple of it, $T=n\tau$, 
the mean energy $\bar{E_1}$ never decreases.

If we denote by $\Phi_1$ the transition map (\ref{eq:transitionmap})
for one period of our periodic system, then the transition map $\Phi_n$ for
an interval of exactly $n$ periods of length $\tau$ is simply a power
of $\Phi_1$,  namely

\be \label{eq:nper}
\Phi_n = \Phi_1^n. 
\ee
If we use units such that $\om_0=\om_1=1$ and $M=1$, then
the mean energy from (\ref{eq:abg}) and (\ref{eq:bare1}) can be 
expressed elegantly as the trace of the product of the transition
matrix $\Phi$ and its transpose $\Phi^T$, namely

\be  \label{eq:trace}
\bar{E_1} = \frac{E_0}{2} (\alpha + \beta) = 
\frac{E_0}{2}  (a^2+b^2+c^2+d^2) = \frac{E_0}{2} {\rm Tr} (\Phi \Phi^T). 
\ee
If  $\Phi=\Phi_n =\Phi_1^n$, then we have to understand the behaviour of
the above expression (\ref{eq:trace}) as a function of the number of
periods $n$. This is obviously dictated by the eigenvalues of
the transition map $\Phi_1$, whose determinant is of course equal to $1$,
and its trace is denoted by $S$, which can be reduced to the diagonal form
by a similarity transformation 

\be \label{eq:similarity}
\Phi_1 = W D W^{-1} 
\ee
where $W$ is the transformation matrix and $D$ is the diagonal matrix with the
eigenvalues  $e_1$ and $e_2=1/e_1$, namely the solutions of the quadratic
equation for $e$,

\be \label{eq:eigen}
e^2 - eS + 1 = 0, \;\;\;  S = {\rm Tr} \Phi_1\;\;\;
e_1 = 1/e_2 = \frac{S}{2} \pm \sqrt{ \left(\frac{S}{2}\right)^2 -1}.
\ee
Therefore $e_1,\; e_2=1/e_1$ are either real reciprocals (if $|S|>2$) 
or complex conjugated  numbers on the unit circle (if $|S|<2$). 
The $n$-th power of $\Phi_1$ can then be written as

\be \label{eq:Phin}
\Phi= \Phi_n = \Phi_1^n  = W D^n W^{-1}.
\ee
Therefore, the matrix elements $a,b,c,d$ of $\Phi_n$ are bounded when
$|S|<2$ and oscillate with $n$, while they increase exponentially when $|S|>2$.
Indeed, if $e_1 >1$, then the asymptotic behaviour of (\ref{eq:trace})
is 

\be \label{eq:bare1per}
\bar{E_1} \approx K E_0 e_1^{2n},
\ee
where $K$ is a constant determined by the matrix $W$
in the transformation (\ref{eq:similarity}),
and the variance at sufficiently large $n$ goes asymptotically as

\be \label{eq:mu2per}
\mu^2 = \overline{(\Delta E_1)^2} \approx \half \bar{E_1}^2 \approx 
 \frac{K^2}{2} E_0^2 e_1^{4n}.
\ee
Therefore, in case when $e_1>1$, we indeed find that the mean energy
and with it the variance of the energy increase exponentially with time
$T= n\tau$, and since $\omega(t)$ is bounded, nothing is conserved.
$e_1$ is determined by $S = {\rm Tr} \Phi_1$, and this in turn is
determined by the specific properties of the system, that is by
the nature of $\om (t)$ on the interval $(0,\tau)$. 

So far we discussed the statistical properties of the energy distribution
$P(E_n)$ in a periodic system. 
Of course, the contour of the initial uniform canonical
ensemble ${\cal K}_0$ is topologically always a circle, it evolves into
the closed curve ${\cal K}_n$ after the $n$-th full period, 
with the preserved, constant, area enclosed by ${\cal K}_n$.
This curve is just rotating and oscillating with $n$
in case $|{\rm Tr}\Phi_1|= |S| <2$. If $ |S|>2$,  the action by 
$\Phi_n=\Phi_1^n$ is exponentially stretching in the direction of
the eigenvector with eigenvalue $e_1>1$, and exponentially contracting
in the direction of the other eigenvector with the eigenvalue
$e_2=1/e_1<1$. Therefore, the energy of the individual initial condition
described by the formula (\ref{eq:em1}) will be exponentially
increasing for {\em any} initial condition $(q_0,p_0)$ (vector
in the phase plane $(q,p)$), except for the case when $(q_0,p_0)$ is
exactly in the direction of the second eigenvector (stable manifold)
corresponding to $e_2=1/e_1 <1$.

\section{General formula for the energy evolution}

In this section we wish to consider an exact expression for the evolution of
the energy distribution by studying a decomposition of one adiabatic
process into several consecutive adiabatic processes.

Let us first mention that the energy distribution $P(E_1)$ evolved from
the original delta-like distribution $\delta (E-E_0)$ is a kind of a Green
function for the energy evolution. Let us denote it by 
$G(E_1;E_0)$.  If we have a distribution of initial
energies $w(E_0)$, such that at each energy $E_0$
the distribution on the energy contour is a uniform canonical
distribution, then the final energy distribution is

\be \label{eq:Green}
P(E_1) = \int G(E_1;E_0) w(E_0) dE_0.
\ee
Thus by knowing $G$, which we call $G$-function,
we can calculate the final energies of 
any family $w(E_0)$ of uniform canonical ensembles of initial conditions.

If the adiabatic process is ideal adiabatic, then the
$G$-function is a delta function,

\be \label{eq:Greenidealad}
G(E_1;E_0) = \delta (E_1 - \ok E_0/\on).
\ee
For ensembles of other types, which are not uniform canonical,
we must go back to our fundamental equation  (\ref{eq:ee1})
and perform the averaging using the distribution in space
$(E_0,\phi)$.

Now suppose that the interval of length $T$ is divided into an arbitrary
number of finite subintervals $(t_j,t_{j+1})$, where $t_0$ is the
beginning of the process (interval) and $t_n$ is the end of the process, 
and $j=0,1,\dots, n-1$.
The behaviour of $\om (t)$ inside each $j$-th subinterval is assumed
to be entirely arbitrary, but the process must be such that
at each integration step $t_j$ the distribution is uniform canonical.
This condition is certainly satisfied if the process is ideal
adiabatic, in general not.
 
It is then obvious that the energy 
$G$-function $G(E;E_0)$ for the complete process divided into
$n$ subintervals is given by the multiple integral

\bea \label{eq:Gcomposition}
G(E;E_0) = \\ \nonumber
\underbrace{\int\dots \int}_{n-1} G_n(E;x_{n-1}) G_{n-1}(x_{n-1};x_{n-2})
\dots G_1(x_1;E_0) dx_{n-1} \dots dx_2 dx_1.
\eea
Indeed, using (\ref{eq:Greenidealad}) we can immediately verify
this equation. 
All moments of the final distribution can be easily calculated
as they are all fully determined by the first moment alone, so what
we need is just the first moment.
From our theory in section 2 we know that in full generality
the first moment of any $G(E;E_0)$ is a linear function of the
initial value $E_0$, namely

\be \label{eq:Gmean}
\bar{E} = \int E G(E;E_0) dE = g E_0,
\ee
where the constant $g= (\alpha+\beta)/2$ is a constant independent
of $E_0$ and is determined by the nature of $\om (t)$ inside the
relevant interval of evolution. We shall call $g$ the $g$-factor of $G$.
For an ideal adiabatic process we know from (\ref{eq:idealadia})
that $g_j=\om_j/\om_{j-1}$.
From equation (\ref{eq:Gcomposition}) we immediately find
the factorization property

\be \label{eq:gfact}
g = g_ng_{n-1} \dots g_2 g_1,\;\;\; \bar{E} = g E_0 = g_n\dots g_2g_1 E_0.
\ee
Obviously, for an ideal adiabatic process where each $g_j=\om_j/\om_{j-1}$,
the above equation is certainly satisfied.

It is possible also to show the converse \cite{RR2006new}: If 
the composition formula (\ref{eq:Gcomposition}) is true for
{\em any} intermediate points of integration $t_j$ and $x_j$,
then the process must be ideal adiabatic, implying that

\be  \label{eq:Gidealadia}
G(E_j;E_{j-1}) = \delta (E_j - \om_j E_{j-1}/\om_{j-1})
\ee
applies for all $j$, and $g_j=\om_j/\om_{j-1}$.
This can be shown by splitting the time interval $(t_0,t_n)$
into infinitesimal subintervals and using a piecewise
constant function to approximate $\om (t)$, and then using
$g_j = \frac{1}{2} (\om_j^2/\om_{j-1}^2 +1)$ from equation
(\ref{eq:jump}) for all $j$, finally evaluating $g$ by
equation (\ref{eq:gfact}), and finding $g=\om_n/\om_0$,
which implies that the process is ideal adiabatic at all
times of the time interval, because $\mu^2=0$.

The composition formula (\ref{eq:Gcomposition}) (factorization
property of the $G$-function) will apply also in nonlinear 
systems, but the relationship between $\bar{E_1}$ and $E_0$ 
is then no longer linear. Therefore using the composition
formula (\ref{eq:Gcomposition}) for infinitesimal intervals,
and approximating $\om (t)$ by piecewise constant or
piecewise linear functions etc. might be of extreme importance
to find new global powerful approximations for $G$-functions and 
their moments.

The theory for nonlinear systems is left open for the future work.

\section{Discussion and conclusions}

Our aim in this work is to study the time evolution of the energy
in a general ({\em no restriction upon $\om (t)$}) 
time-dependent 1D harmonic oscillator in a rigorous way,
and then also to calculate the statistical properties of the final
energy distribution $P(E_1)$ after some time of lentgh $T$,
if at the beginning we have a uniform canonical distribution of
initial conditions at constant energy $E_0$. 
We are able to calculate rigorously all  the moments of $P(E_1)$.
Odd moments are exactly zero, the even moments are powers of
the second moment and the second moment $\mu^2$ is a function
of the first moment. Therefore everything is determined
by the first moment $\bar{E_1}$ and the variance $\mu^2$,
which is a function of $\bar{E_1}$. Thus the energy distribution
function $P(E_1)$ is universal in the sense that it does not
depend on any other properties of $\om (t)$.
(Mathematically speaking, the distribution function  $P(E_1)$ is a two
parameter family parametrized by $\bar{E_1}$ and $\mu^2$, but
in physics these two parameters are connected through
the formula (\ref{eq:sigma}). Of course, the shape of the
distribution function $P(E_1)$ does not change if $\bar{E_1}$ is
translationally shifted, and thus for the mathematical purposes we can
set $\bar{E_1}=0$.)
In this analysis we clearly see when the adiabatic
invariant $I(t)=E(t)/\om (t)$ is conserved or not. 
In the (ideal) adiabatic limit $T\rightarrow \infty$ it is conserved,
the variance $\mu^2$ is zero and $E_1=\bar{E_1}=\om_1 E_0/\om_0$.
If it is not conserved exactly, when $T$ is finite, 
we find $\mu^2 > 0$, and it can be calculated using a WKB method
analytically in a closed form, which is a major achievement of this work.
We have also studied three specific solvable models and have 
demonstrated the power of the WKB expansion, where already the
leading WKB term usually very well describes the asymptotic behaviour
of $\mu^2$  when $\ep =1/T$ goes to zero. We also show what happens
if $\om (t)$ is smooth and of class ${\cal C}^m$, having $m$ continuous
derivatives, calculating and showing that $\mu^2$ oscillates 
as $\ep$ goes to zero, but in the mean vanishes as $\propto \ep^{2(m+1)}$.
If $\om (t)$ is analytic, thus it also is of class ${\cal C}^{\infty}$,
it is known from the literature that $\mu^2$ must decay exponentially
$\propto \exp (const/\ep)$. If $\om (t)$ is periodic, $\bar{E_1}$ can
grow exponentially, and so does the variance $\mu^2$, in which case
$I(t) = E(t)/\om (t)$ is not conserved, but we can describe the system.
We have introduced the so-called $G$-function, which is a kind of
a Green function for the evolution of the energy and derived a
composition formula for it when the interval of evolution
is decomposed into a finite number of subintervals, for which the
corresponding $G_j$-function is known for all subintervals $j$ and 
is uniform canonical there.
This formula applies also to nonlinear systems and might be
a good starting basis to describe them. The 
theory for nonlinear systems remains open and is a subject
of the current research \cite{RR2006new}.

\vspace{0.3in}


\section*{Acknowledgements}

This work was supported by the Ministry of Higher Education, Science and 
Technology of the Republic of Slovenia, by the Nova Kreditna Banka Maribor 
and by TELEKOM Slovenije. M.R. gratefully acknowledges
the hospitality of Dr. Takahisa Harayama at ATR, Seika-cho, Japan,
during his research visit there.

\end{document}